\documentclass[aps,pra,twocolumn,showpacs, superscriptaddress]{revtex4-1}

\usepackage{graphicx}
\usepackage{color}
\usepackage[usenames,dvipsnames]{xcolor}

\newcommand{\nc}{\newcommand}
\nc{\eq}{\begin{equation}}
\nc{\eeq}{\end{equation}}
\nc{\eqa}{\begin{eqnarray}}
\nc{\eeqa}{\end{eqnarray}}
\nc{\ar}{\begin{array}}
\nc{\ear}{\end{array}}
\nc{\bfig}{\begin{figure}}
\nc{\efig}{\end{figure}}
\nc{\dg}{\dagger}
\nc{\eps}{\frac{\epsilon}{2}}
\nc{\juuri}{\sqrt{\Omega^2+(\eps)^2}}
\nc{\sx}{\sigma_x}
\nc{\sy}{\sigma_y}
\nc{\sz}{\sigma_z}
\nc{\spl}{\sigma_+}
\nc{\sm}{\sigma_-}
\nc{\Sx}{\bar{\sigma}_x}
\nc{\Sy}{\bar{\sigma}_y}
\nc{\Sz}{\bar{\sigma}_z}
\nc{\Spl}{\bar{\sigma}_+}
\nc{\Sm}{\bar{\sigma}_-}
\nc{\nn}{\nonumber}
\nc{\noi}{\noindent}
\nc{\omt}{\tilde{\omega}}
\nc{\Somt}{S(\omt)}
\nc{\Somtd}{S^{\dg}(\omt)}
\nc{\got}{\gamma_{\omega}(t)}
\nc{\gmot}{\gamma_{-\omega}(t)}
\nc{\po}{\mathcal{P}}
\nc{\qo}{\mathcal{Q}}
\nc{\adg}{a^{\dg}}
\nc{\gammat}{\tilde{\gamma}}
\nc{\Q}{$\mathcal{Q }$}
\nc{\C}{$\mathcal{C }$}
\nc{\kvec}{\mathbf{k}}

\def\ket#1{\mathinner{|{#1}\rangle}}

\begin{document}


\title{Two-qubit non-Markovianity induced by a common environment}



\author{C. Addis}
\email{ca99@hw.ac.uk}
\affiliation{Scottish Universities Physics Alliance, Engineering and Physical Sciences, EPS/Physics, Heriot-Watt University, Edinburgh, EH14 4AS, UK}
\author{P. Haikka}
\email[]{pmehai@utu.fi}
\affiliation{Turku Center for Quantum Physics, Department of Physics and Astronomy, University of Turku, FIN-20014 Turku, Finland}
\author{S. McEndoo}
\affiliation{SUPA, EPS/Physics, Heriot-Watt University, Edinburgh, EH14 4AS, UK}
\author{C. Macchiavello}
\affiliation{Dipartimento di Fisica and INFN-Sezione di Pavia, Via Bassi 6, 27100 Pavia, Italy}
\author{S. Maniscalco} 
\email[]{s.maniscalco@hw.ac.uk} 
\affiliation{SUPA, EPS/Physics, Heriot-Watt University, Edinburgh, EH14 4AS, UK}
\affiliation{Turku Center for Quantum Physics, Department of Physics and Astronomy, University of Turku, FIN-20014 Turku, Finland}


\date{\today}

\begin{abstract}
We study non-Markovianity as backflow of information in two-qubit systems. We consider a setting where, by changing the distance between the qubits, one can interpolate between independent reservoir and common reservoir scenarios. We demonstrate that non-Markovianity can be induced by the common reservoir and  single out the physical origin of this phenomenon. We show that two-qubit non-Markovianity coincides with instances of non-divisibility of the corresponding dynamical map, and we discuss the pair of states maximizing information flowback. We also discuss the issue of additivity for the measure we use and in doing so, give an indication of its usefulness as a resource for multipartite quantum systems. 
\end{abstract}


\maketitle


\section{Introduction}

Exploiting quantum correlations, such as entanglement, as a resource for quantum information revolves around the major obstacle of preserving them from the detrimental interaction with the environment \cite{bp, weiss}. Efficient schemes for the protection of correlations generally require detailed knowledge of the microscopic processes that induce environmental noise. To this aim we consider a simple bipartite system of two spatially separated qubits immersed in a non-Markovian structured environment. In such an environment memory effects exist as a result of complex system-environment coupling, and consequently quantum properties can be temporarily restored. Motivated by the future prospect of scalable quantum devices requiring multipartite systems, and in view of the fact that non-Markovianity is a resource for quantum technologies \cite{Bogna}, we study how environment-mediated interactions enhance memory-keeping properties and affect the behavior of non-Markovianity in two-qubit channels. Our main focus is on correlated noise induced by the common environment. By changing the separation between the qubits we are able to interpolate between local and common reservoir scenarios and explicitly study the effect of environment-mediated correlations on the degree of non-Markovianity of the two-qubit dynamical process.

In order to systematically investigate the ability of an open system to regain quantum properties, it is important to unambiguously define and quantify non-Markovian dynamical behavior. Here we consider two measures of non-Markovianity recently introduced by Breuer, Laine, and Piilo (BLP) \cite{BLP} and Rivas, Huelga, and Plenio (RHP) \cite{RHP}. These measures are based on similar but subtly different definitions of non-Markovianity, i.e., backflow of information from the environment to the system and non-divisibility of the dynamical process, respectively. The definitions agree for some dynamical processes; however, this is generally not the case  \cite{ref1, ref2, ref3}. The origin and extent of differences of these measures are active areas of research, albeit with most current studies focusing on single-qubit dynamical processes. The present study, comparing and contrasting the measures for a two-qubit system, is therefore also of fundamental interest.

Non-Markovianity measures characterize and quantify certain properties of quantum dynamical maps and, therefore, do not depend on the initial state of the open quantum system. In practice this means that they are defined via optimization over all possible initial states. As a consequence they are increasingly difficult to calculate for multi-qubit systems. Moreover, these measures are in general not additive; hence, it is not possible to reduce the calculation of non-Markovianity of an $n$-qubit channel to the one of a single-qubit channel even when the qubits are subjected to identical local uncorrelated reservoirs \cite{Fanchini2013}. All previous studies on non-Markovianity as information flow in two-qubit systems focus on the simpler case of local environments \cite{Fanchini2013, Shao2010, Two_Q, elsitwoqubit}. The model considered here studies BLP and RHP non-Markovianity in a common environment scenario in order to demonstrate how environment-mediated correlations create non-Markovian dynamics via correlated noise. 

This article is organized as follows. In Sec. II we introduce our model and the master equation describing the dynamical evolution of the system. In Secs. III and IV we study the two non-Markovianity measures mentioned above in relation to relevant system parameters,commenting on the optimizing pair and the additivity property of the BLP measure. Finally in Sec. V we summarize our results and present our conclusions.

\section{The Model}

We consider a system of two qubits, spatially separated by distance $2D$ and dephasing under the influence of a bosonic reservoir with a non-trivial spectral density function. The model was introduced originally in Ref. \cite{kalleantti, reina}; however, we will later specify our calculations on a physically realistic system with a more complex spectral density function \cite{originalmodel}. When $D \rightarrow \infty$ the dynamical map factorizes into the product of the dynamical maps of the individual qubits, $\Phi(t) = \Phi_A(t) \otimes \Phi_B(t) $, where subscript $A$ refers to the first qubit and $B$ refers to the second one. For finite distances $D$, however, this is not the case and the qubits are subjected to correlated noise, i.e., to environment-mediated interactions. On the one hand, the conditions for this system to induce non-Markovian dynamics on a \emph{single} qubit have been recently studied in Refs. \cite{us, us2}. On the other hand, a link between the dynamical evolution of the qubits and their spatial separation has been considered in Refs. \cite{doll, doll2}, even if not in connection with reservoir memory effects. Our goal is to study the interplay between the environmental spectrum and the spatial separation in creating non-Markovian dynamics for this two-qubit model, focusing especially on the ability of correlated quantum noise to induce a backflow of information.

We have derived the master equation describing the dynamics of the two qubits starting from the exact analytical expression of the density matrix of two qubits at all times given in Refs. \cite{kalleantti,reina}. For the model here studied it is possible to solve the dynamics of the total closed system exactly and obtain the reduced system dynamics by taking the partial trace. Following straightforward calculations, the master equation can be recast in the following Lindblad-type form with time-dependent coefficients.
\begin{widetext}
\begin{eqnarray}\label{masterequation}
\frac{d\rho}{dt}&=&\frac{\gamma_1(t)-\gamma_2(t)}{2}\left[(\sz^A-\sz^B)\rho(\sz^A-\sz^B)-\frac{1}{2}\left\{(\sz^A-\sz^B)(\sz^A-\sz^B),\rho\right\}\right]\nn\\
&+&\frac{\gamma_1(t)+\gamma_2(t)}{2}\left[(\sz^A+\sz^B)\rho(\sz^A+\sz^B)-\frac{1}{2}\left\{(\sz^A+\sz^B)(\sz^A+\sz^B),\rho\right\}\right],
\end{eqnarray}
\end{widetext}
where the decay rates are given as the sum and the difference of
\eq
\gamma_{1}(t)=\frac{1}{2}\frac{d\Gamma_0(t)}{dt}, \quad\gamma_{2}(t)=\frac{1}{4}\frac{d\delta(t)}{dt}.
\eeq
Here $\Gamma_0(t)$ is the decoherence function describing individual dephasing of each qubit and $\delta(t)$ encapsulates the corrections to the single qubit dephasing arising from environment-mediated interactions. In the limit of the qubit separation going to infinity the latter vanishes and the master equation reduces to a sum of two Lindblad-type terms describing the individual dephasing of each qubit with decay rate $\gamma_1(t)$. We note that the single-qubit decoherence function $\Gamma_0(t)$ and the correction term $\delta(t)$ are connected to the \emph{collective decoherence functions},
\eq
\Gamma_\pm(t)=2\Gamma_0(t)\pm\delta(t),
\eeq
of Ref. \cite{kalleantti}. Physically $\Gamma_+(t)$ and $\Gamma_-(t)$ describe the decay of the super- and sub-decoherent two-qubit states, $\ket{\Psi_\pm}=\frac{1}{\sqrt{2}}(\ket{10}\pm\ket{01})$ and $\ket{\Phi_\pm}=\frac{1}{\sqrt{2}}(\ket{00}\pm\ket{11})$, respectively. Decay of the super-decoherent state is enhanced by environment-mediated interaction, whereas the decay of the sub-decoherent state is suppressed by it. It should be noted that the explicit form and physical meaning of the decay rates $\Gamma_{\pm}(t)$ appearing in the two-qubit channel remain general while their functional form is dependent on the specific model studied. \\
Initially the two qubits dephase individually as if coupled to independent environments. At some later time, if the qubits are sufficiently close together, they are able to ``see" each other via correlations arising through an interaction mediated between them by the environment, formally contained in $\delta(t)$. This interaction becomes increasingly weak as the separation between the qubits increases, effectively approaching zero at sufficiently large distances \cite{originalmodel}.  This model is therefore particularly interesting because it allows insight into these two possible physical scenarios through simply changing the separation between the qubits.\ In the following sections we study the dynamics of the decay rates, connecting them to two prevailing definitions of non-Markovianity.

\section{Decay rates and divisibility}

Useful information on the dynamical properties of the two-qubit system is contained in the specific time dependence of the decoherence rates appearing in the master equation Eq. (\ref{masterequation}). A completely positive, trace preserving (CPT) map $\Phi(t,0): \rho(0)\mapsto\rho(t)$ is called divisible if $\Phi(t,0)=\Phi(t,t')\Phi(t',0)$, where $\Phi(t,t')$ and $\Phi(t',0)$ are CPT maps $\forall{t'}$. For master equations in the diagonal form, such as the one of Eq. (\ref{masterequation}), the divisibility property is reflected in the coefficients of the master equation. Divisibility, in turn, is crucially related to non-Markovianity. For all known non-Markovianity measures ${\cal N}$ a finite value ${\cal N}\neq 0$ implies that the dynamical map is non-divisible. In particular, the RHP measure explicitly quantifies non-divisibility of the dynamical map, ${\cal N}_{\text{RHP}}\neq 0$ if and only if the dynamical map is non-divisible \cite{RHP}.

For time-local master equations in Lindblad-type form the map is divisible whenever all time-dependent decoherence rates, i.e., $\gamma_1(t) \pm \gamma_2(t)$ in the case of our model, are positive at all times. Conversely, if at least one of the time-dependent coefficients temporarily takes negative values, then the dynamical map is not divisible.

Since the negativity of the decay rates depends crucially on the spectral density function characterizing the properties of the environment and the qubit-environment coupling, in the following, for the sake of concreteness, we will focus on a specific experimentally realizable physical system. The set-up we consider comprises two impurity atoms dephasing due to an interaction with an ultracold bosonic rubidium gas in a Bose-Einstein condensed (BEC) state, introduced in detail Appendix. Each atom is trapped in double well potential, forming a qubit system where the two qubit states are represented by the occupation of the atom in either the left or the right well. The two impurity atoms are separated by distance $2D\ge 8L$, with $2L$ being the distance between the two minima for the double-well potential, i.e. the ``size" of the qubit. It has been shown in Refs. \cite{us, us2} that this qubit system is particularly sensitive to non-Markovian effects. The properties of the bosonic gas can be controlled by modifying its trapping potential (in this way one may change the effective dimensionality of the BEC) and/or changing the boson-boson scattering length $a_B$ from it's natural value $a_{Rb}$ via Feshbach resonances (allowing interpolation from a free to an interacting background gas). The dynamics of a single impurity immersed in the BEC gas was studied in Ref. \cite{us}, where it was reported that the dynamics of the single dephasing qubit is sensitive to the scattering length $a_B$ of the BEC. A critical value of the scattering length, denoted $a_B^{crit}$ and depending on the dimensionality of the BEC, was found to specify where the crossover between Markovian and non-Markovian behavior occurred. If the background gas is non-interacting or very weakly interacting, i.e., $0\le{a_B}\le{a_B^{crit}}$, the qubit dynamics is Markovian. If the background gas has stronger interactions, i.e., $a_B\textgreater{a_B^{crit}}$, qubit dynamics has signatures of reservoir memory effects.

We now study the non-divisibility of the two-qubit dynamical map, signalled by temporarily negative decay rates. Choosing the BEC scattering length and the distance between the two qubits appropriately, we can consider the cases when the single qubit dynamics is Markovian or non-Markovian and the cases when the qubits are effectively in local or common environments. 

\begin{figure}[h]
\includegraphics[width=0.5\textwidth]{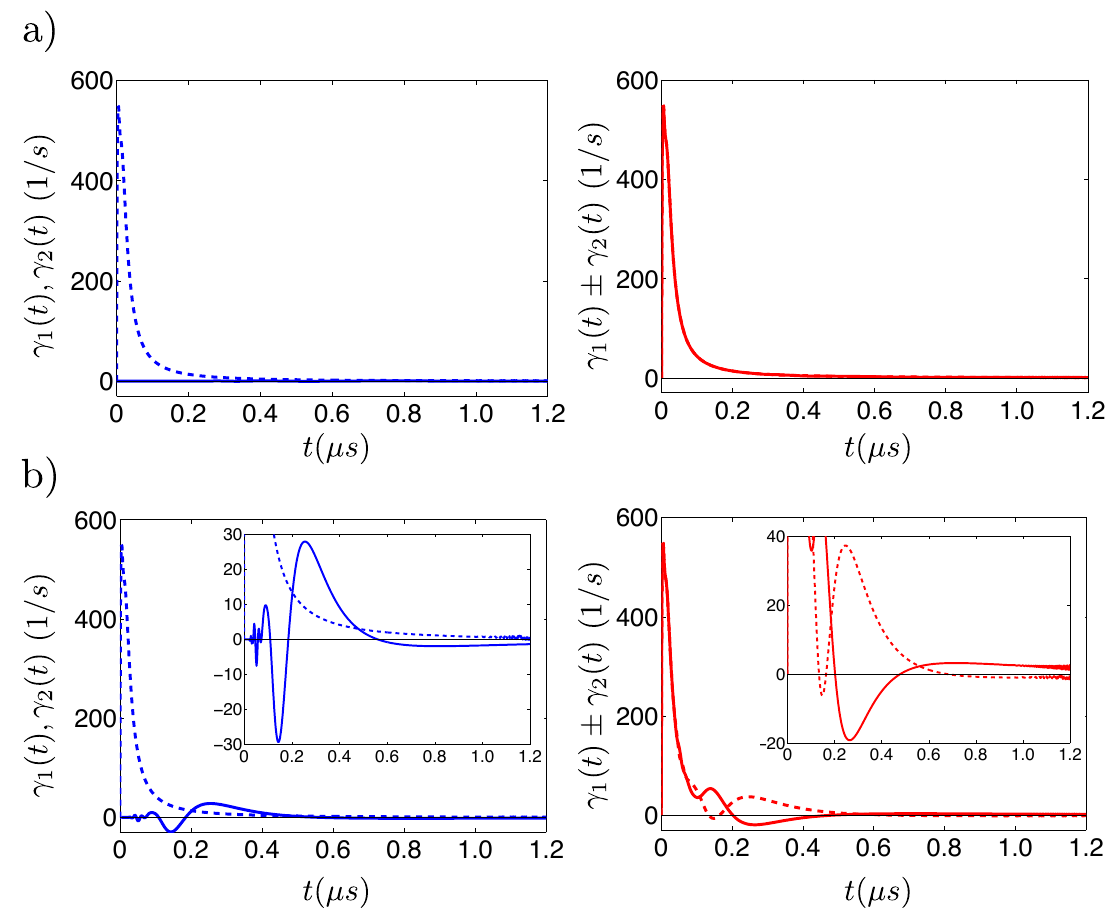} 
\caption{(Color online) a) The independent reservoir, $D=200L$. b) The common reservoir case, $D=4L$. Time-dependent coefficients $\gamma_1(t)$ (blue dashed line) and $\gamma_2(t)$ (blue solid line) are shown in the first column, with $\gamma_1(t)+\gamma_2(t)$ (red dashed line) and $\gamma_1(t)-\gamma_2(t)$ (red solid line) in the second column. The scattering length is chosen to be $0.02a_{Rb}<a_B^{crit}$ to ensure the one qubit channel is Markovian for all times $t$ and $a_{Rb}$ is the natural scattering length of a rubidium condensate; see the Appendix. The insets zoom in on small values of the decay rates.} 
\label{decayrates_2}
\end{figure} 
We begin by looking at the case in which $D/L\gg 1$, i.e., the qubits are subjected to uncorrelated noise, and values of the scattering length such that each one-qubit dynamical map is Markovian, i.e., $\gamma_1(t)$ is always positive (see Fig. 1, upper row). We note that, while $\gamma_2(t)\approx 0$ at all times as expected, $\gamma_1(t)$ increases initially but, after a certain time, decreases to zero, meaning that the dephasing of the individual qubits stops.

Keeping all other parameters fixed, we now decrease the distance between the qubits, and we see that, after a short time interval, the cross-talk term $\gamma_2(t)$ begins taking negative values (see Fig. 1, lower row). More precisely, the maximum negative value of $\gamma_2(t)$ is reached when $\gamma_1(t)$ is already significantly small. As a consequence the decoherence rate $\gamma_1(t) + \gamma_2(t)$ takes negative values, indicating that the two-qubit dynamical map is non-divisible. This feature is not specific to the values of parameters chosen in Fig. 1, but occurs for any initial values of $a_B$ such that, when  $D/L\gg 1$, the map is Markovian. Therefore we can uniquely link the presence of correlated noise to the violation of divisibility and, hence, to a non-zero value of the RHP non-Markovianity measure.

To conclude, the presence of the ``cross-talk" term $\gamma_2(t)$ is sufficient to break divisibility in the combined system regardless of the fact that the individual qubit dynamics is divisible. More generally, whenever the qubits are subjected to Markovian local dynamics, i.e., $\gamma_1(t)$ (a term independent of qubit separation) is always positive, non-divisible non-local dynamics of the two qubits can result only from the action of correlated noise. 

In order to gain further insight in the microscopic physical processes of the composite system, we will look in the next section to the behavior of information flux. Our aim is to see whether the cross-talk term acts as the reservoir memory, i.e., if it is responsible for backflow of information. To answer this question we need to study the BLP non-Markovianity measure, which, as we mentioned in the Introduction, does not always coincide with the RHP measure.

\section{Information flux and Non-Markovianity Measure}

The BLP measure of non-Markovianity is based on the behavior of information flux between system and environment \cite{BLP}. According to this definition, a system is non-Markovian when part of the information that was lost into the environment due to decoherence and/or dissipation is temporarily restored in the system. To quantify this process one can use the trace distance  between two states as a measure of their distinguishability and equate decrease of distinguishability with loss of information about the system. Hence, the rate of change in trace distance $D(\rho_1(t),\rho_2(t))=\frac{1}{2}Tr|\rho_1(t)-\rho_2(t)|$, where $\rho_{1,2}(t)=\Phi(t,0)\rho_{1,2}(0)$ are two initial states evolving under the dynamical map $\Phi(t,0)$, can be interpreted as an information flux $\sigma$,
\eq \label{2}
\sigma(t,\rho_{1,2}(0))=\frac{d}{dt}D(\rho_1(t),\rho_2(t)).
\eeq \\
A decrease in the trace distance represents information flowing from the system into the environment and an increase represents a back flow of information from the environment to the system. The total increase of distinguishability over the whole time evolution quantifies the total amount of information flowing from the environment back into the system and therefore is defined as a measure for non-Markovianity:
\eq  \label{N}
\quad\mathcal{N}_{BLP}=\max_{\rho_1,\rho_2}\int_{\sigma>0}ds\, \sigma(s).
\eeq 
The time integration is extended over all time intervals for which $\sigma$ is positive, and the maximum is taken over all pairs of initial states, $\rho_{1,2}$. Therefore the non-Markovianity is defined as the maximal amount of information that the system can possibly recover from its environment. A dynamical map which is divisible always describes Markovian dynamics in terms of information flow; however, the converse is generally not always true \cite{ref1, ref2, ref3}.

The maximization in Eq. (\ref{N}) complicates the calculation of the BLP measure of non-Markovianity. In Ref. \cite{anttisteffen} the mathematical and physical properties of the optimal pairs corresponding to the maximal backflow of information from the environment to the system are characterized, simplifying the maximization, but finding the optimal pair for a two-qubit dynamical process still remains a difficult task. To this aim, we calculate the measure by randomly generating pairs of states and classifying them into pure, mixed, separable, and maximally entangled states, including the Bell states. 

\subsection{Non-Markovianity Dependence on Qubit Separation}

\begin{figure*}
\includegraphics[width=1.05\textwidth]{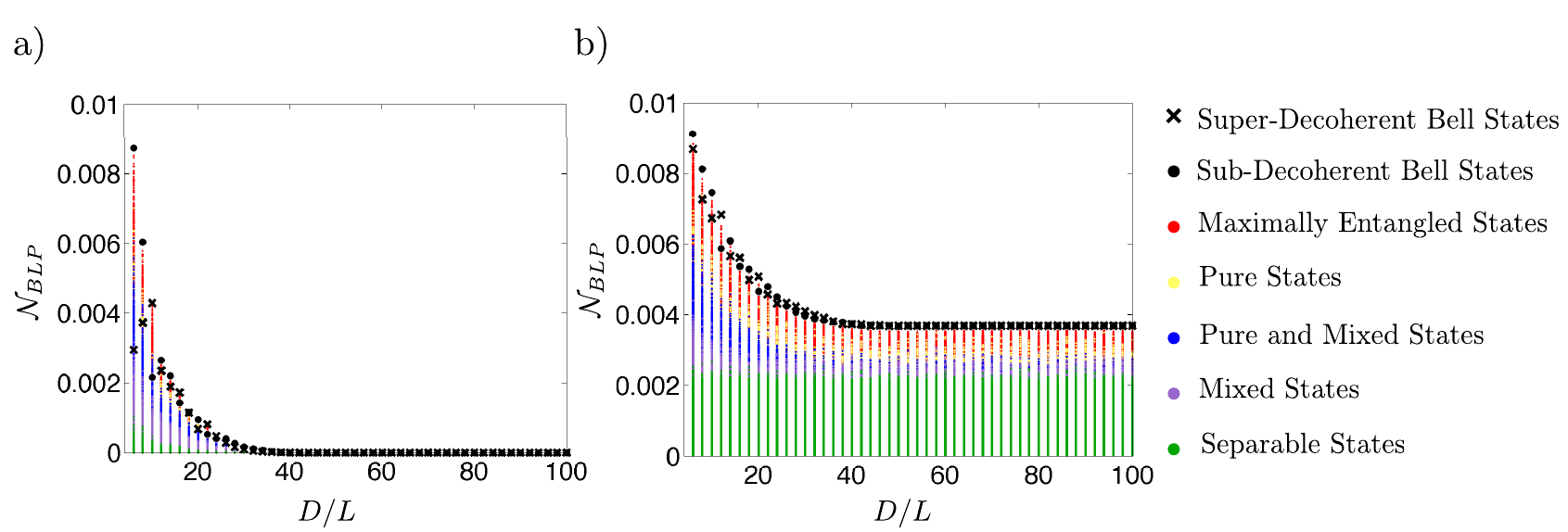} 
\caption{(Color online) Non-Markovianity Measure $\mathcal{N}_{\text{BLP}}$ (the largest value in each column of the plot) as a function of distance $D/L$ between qubits, for scattering lengths a) $0.02a_B^{Rb}<a_B^{crit}$ (locally Markovian dynamics) and b) $0.50a_B^{Rb}>a_B^{crit}$ (locally non-Markovian dynamics). The graph shows 20000 randomly drawn pairs of initial states. The pairs are classified into different categories and color coded accordingly as elaborated in the key on the right. The layers of dense points are given in order (starting from the bottom): separable, mixed, pure and mixed, pure states and maximally entangled states.}
\label{NM1}
\end{figure*}

\begin{figure*}
\includegraphics[width=1\textwidth]{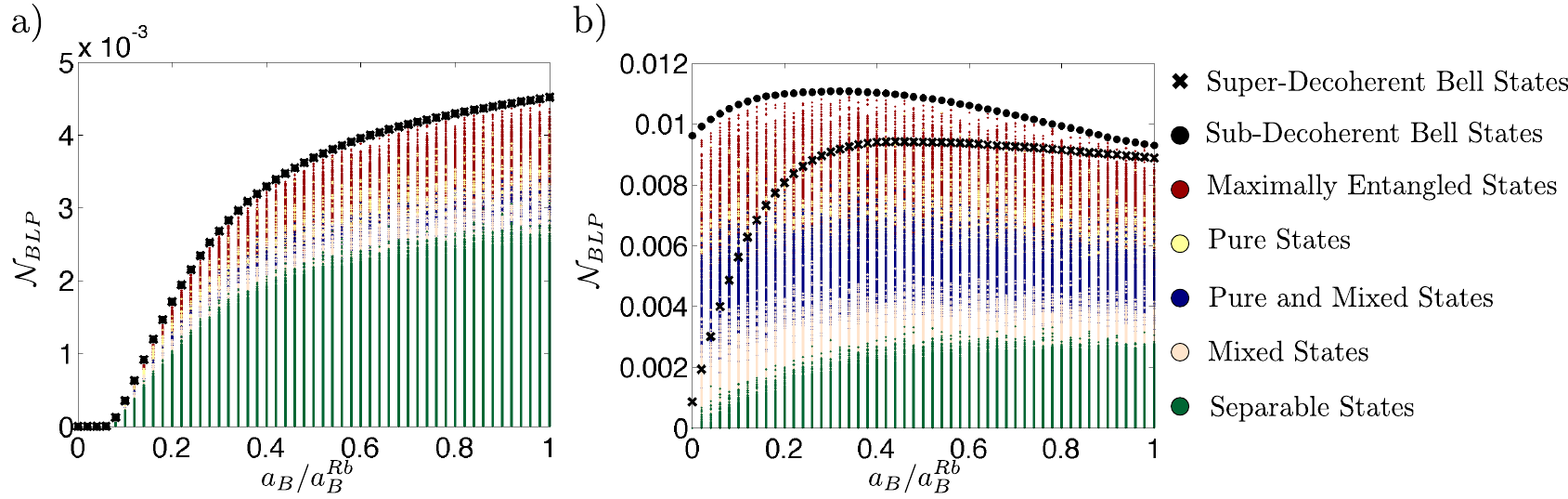} 
\caption{(Color online) Non-Markovianity Measure $\mathcal{N}_{\text{BLP}}$ (the largest value in each column of the plot) as a function of scattering length $a_B$ between qubits for separation a) $D=200L$ (independent reservoir) and b) $D=4L$ (common reservoir). The graph shows 20000 randomly drawn pairs of initial states. The pairs are classified into different categories and color coded accordingly as elaborated in the key on the right. The layers of dense points are given in order (starting from the bottom): separable, mixed, pure and mixed, pure states and maximally entangled states.}
\label{NM2}
\end{figure*}

We will now study how $\mathcal{N}_{BLP}$ changes with the distance between the qubits. We first consider the case of a weakly interacting BEC reservoir inducing local Markovian dephasing of the single qubits (Fig.~\ref{NM1}, left plot). The plot clearly shows that the non-Markovianity measure decreases as the qubits become further and further apart and vanishes for a long enough spatial separation between the qubits. This is consistent with the previous results (see Fig.~\ref{decayrates_2}, first row): when the qubits are very far apart, $D/L\gtrsim30$, they effectively interact with independent environments, and since each qubit individually dephases in a Markovian way, the combined dynamics is also Markovian. For shorter qubit separations, the measure is non-zero and therefore the dynamics is not only indivisible (see Fig.~\ref{decayrates_2}, second row), but also non-Markovian in terms of back flow of information: more precisely, $\mathcal{N}_{BLP} \neq 0$ whenever $\mathcal{N}_{RHP} \neq 0$. We can therefore identify the reservoir memory inducing the information backflow with the correlated noise and, in particular, with the cross-talk term.

When the BEC interactions are stronger, i.e. for larger values of the scattering length $a_B$, the local dephasing is always non-Markovian as the back action of the reservoir on the system is non-negligible also when the qubits are far apart. In this case the BLP measure tends to a constant non-zero value at the point where environment-mediated interactions vanish ($D/L \approx 30$). The value of non-Markovianity saturates and $\mathcal{N}_{BLP}$ does not change for greater distances between the qubits.

\subsection{Non-Markovianity Dependence on Scattering Length}

In order to fully characterize the behavior of information flux, we now study how this quantity changes when, for fixed distances, we vary the scattering length of the ultracold reservoir gas. It is worth mentioning that, while a change in the distance between the qubits can be seen as a change in the effective cutoff frequency of the reservoir spectrum, changing $a_B$ modifies the form of the spectrum itself and, in particular, its Ohmicity character \cite{us}. It has been shown that for purely dephasing qubits the form of the spectrum, and more precisely its low-frequency component, has a strong effect on the presence of information backflow \cite{discord}. Therefore we expect $ \mathcal{N}_{BLP}$ to be sensitive to changes in $a_B$, as can be already seen from Fig. 2.

In Fig.~\ref{NM2} we plot the BLP measure as a function of the scattering length $a_B$ for two values of qubits separations $D/L=200$ and 4, representing an independent (uncorrelated noise) and a common (correlated noise) environment , respectively. For $D/L=200$, if $a_B>a_B^{crit}$, we expect non-Markovian behavior originating purely from the dynamics of the single-qubit. There exists a cross-over between Markovian and non-Markovian processes, as found in Ref.\cite{us}, due to the single qubit dynamical dependence on the scattering length. Only in the presence of environment-mediated interactions between the two qubits (Fig.~\ref{NM2} , $D/L=4$), the combined system can exhibit non-Markovian dynamics for $a_B<a_B^{crit}$. Indeed, we see that even for very weakly interacting gases, we can have $\mathcal{N}_{BLP} \neq 0$. Hence, the presence of correlated noise makes the system non-Markovian for all values of the scattering length. 

Generally, for $a_B>a_B^{crit}$, both single-qubit non-Markovianity and environment-mediated non-Markovianity contribute to the total values of $\mathcal{N}_{BLP}$ in a non-trivial way. While the single-qubit local non-Markovianity, governed by the negativity of $\gamma_1(t)$, monotonically increases with $a_B$ after the threshold value corresponding to Markovian to non-Markovian crossover, non-Markovianity of the composite two-qubit system in the presence of correlated noise initially increases, reaches a maximum value, and then decreases with $a_B$  (Fig.~\ref{NM2}, $D/L=200$). This can be traced back to the behavior of the time-dependent coefficients. In particular, for increasing scattering length, the environment-mediated interactions become weaker [$\gamma_2(t)$ is smaller] and hence the contribution of this term to $\mathcal{N}_{BLP}$ decreases.

\subsection{Optimizing pair}

\begin{figure*}[htp]
\centering
\includegraphics[width=0.6\textwidth]{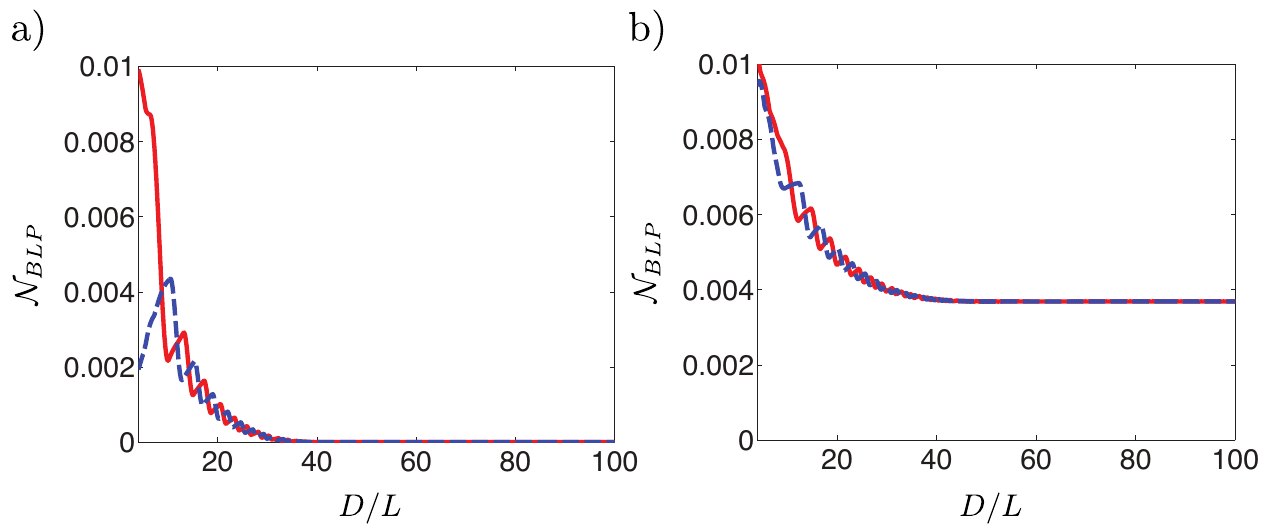}
\caption{Maximising States: Non-Markovianity measure $\mathcal{N}_{\text{BLP}}=\max\{\mathcal{N}_\Phi,\mathcal{N}_\Psi\}$ (the highest value at each instance) as a function of distance between qubits, $D$, for scattering lengths a) $0.02a_B^{Rb}<a_B^{crit}$ (locally Markovian dynamics) and b) $0.50a_B^{Rb}>a_B^{crit}$ (locally non-Markovian dynamics). Red solid line represents $\mathcal{N}_\Phi$ and the blue dashed line represents $\mathcal{N}_\Psi$.}
\label{optimizing}
\end{figure*}

In this section we examine the optimization procedure on which the BLP measure is based by studying numerically the pair of states for which the backflow of information is maximized. This is arguably the most complicated system for which the optimizing pair has been studied in full detail. Our analysis shows that the optimizing pair is extremely sensitive to even small variations in the values of the open system parameters. More specifically, the pair optimizing the BLP measure alternates between two different pairs of states when changing both the scattering length and the distance between the qubits. This underlines the practical difficulties in using the BLP measure for increasing number of qubits.

We have compelling numerical evidence that there are two candidates for the pair maximizing the measure. The first pair $(\Phi^{+},\Phi^{-})$ corresponds to the sub-decoherent Bell states [decohering with rate $\Gamma_-(t)$], while the second pair $(\Psi^{+},\Psi^{-})$ comprises the super-decoherent Bell states [decohering with rate $\Gamma_+(t)$]. The analytical expression for the BLP measure is then
\eq \label{analyticalBLP}
\mathcal{N}_{\text{BLP}}=\max\{\mathcal{N}_\Phi,\mathcal{N}_\Psi\},
\eeq
where $\mathcal{N}_\Phi=\sum_{i}e^{-\Gamma_-(b_i)}-e^{-\Gamma_-(a_i)}$ is the measure if the sub-decoherent Bell states form the maximizing pair and $\mathcal{N}_\Psi=\sum_{i}e^{-\Gamma_+(b_i)}-e^{-\Gamma_+(a_i)}$ is the measure if the super-decoherent Bell states form the maximizing pair. Time intervals $t\in[a_i,b_i]$ indicate the periods of information backflow, manifested as $d\Gamma_\pm(t)/dt\sim\gamma_1(t)\pm\gamma_2(t)<0$.

We demonstrate the sensitivity of the maximizing pair on the model parameters in Fig.~\ref{optimizing} by plotting $\mathcal{N}_\Phi$ and $\mathcal{N}_\Psi$ as a function of the qubit separation. When the qubits share a common environment, corresponding to $D/L\lesssim30$, the pair maximizing the measure depends strongly on the distance between the two qubits, alternating between the two sub-decoherent and the super-decoherent pairs. This can be traced back to the dynamical behavior of $\gamma_2(t)$ and its absolute value in relation to the value of $\gamma_1(t)$. Due to the complex evolution of $\gamma_2(t)$ the relative values of the two decoherence functions $\Gamma_\pm(t)$ oscillate with increasing distance, which is directly reflected in the relative values of  $\mathcal{N}_\Phi$ and $\mathcal{N}_\Psi$. For large qubit separations, the decoherence factors converge in the absence of environment-mediated interactions, $\Gamma_\pm(t)\rightarrow2\Gamma_0(t)$. Consequently for $D/L\gtrsim30$, the differences between $\mathcal{N}_\Phi$ and $\mathcal{N}_\Psi$ diminish until in the case of effectively local environments both pairs maximize the measure.

Equation (\ref{analyticalBLP}) also demonstrates the qualitative agreement of the BLP and RHP measures of non-Markovianity. Recall that the latter is non-zero if either of the decay rates $\gamma_1(t)\pm\gamma_2(t)$ takes temporarily negative values. This immediately implies that either $\mathcal{N}_\Phi$ or $\mathcal{N}_\Psi$ is non-zero, therefore proving that for the model considered here non-divisibility of the two-qubit map exactly coincides with information backflow.

\subsection{Additivity issues for the non-Markovianity measure}

\begin{figure*}[htp]
\centering
\includegraphics[width=0.85\textwidth]{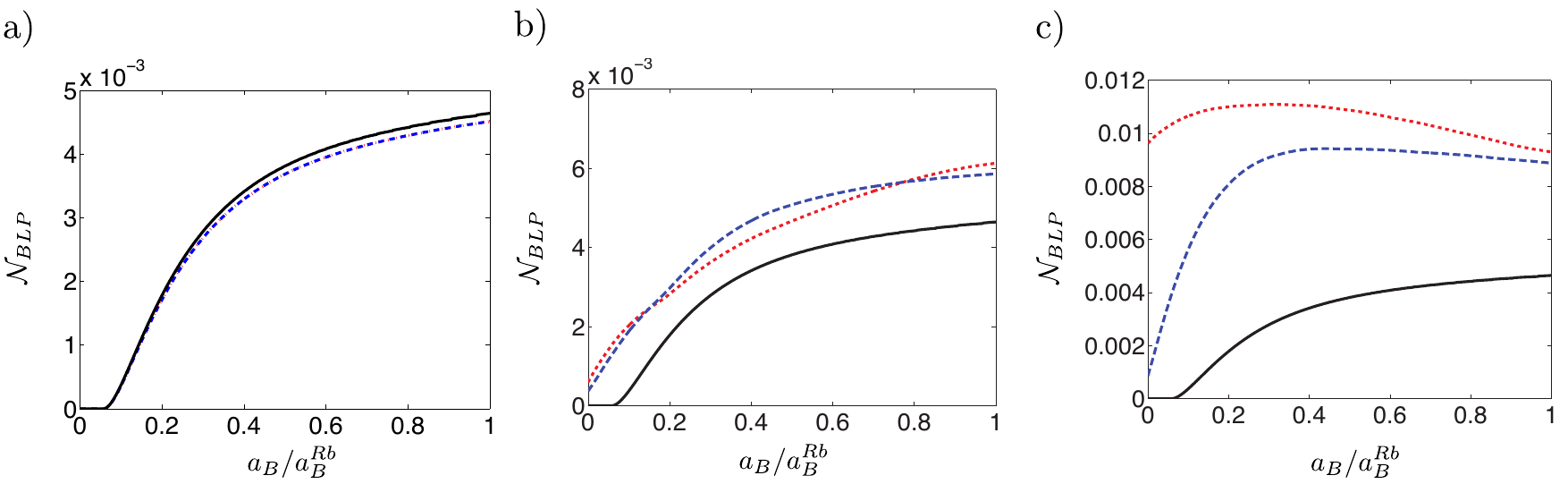}
\caption{Additivity of the non-Markovianity measure: $\mathcal{N}_\Phi$ (red solid line), $\mathcal{N}_\Psi$ (blue dashed line) and 2$\mathcal{N}_1$ (black solid line) as a function of scattering length $a_B$, for qubit separations of a) $D=200L$, b) $D=20L$ and c) $D=4L$. Recall that $\mathcal{N}_2=\mathcal{N}_{\text{BLP}}=\max\{\mathcal{N}_\Phi,\mathcal{N}_\Psi\}$.}
\label{nM1}
\end{figure*}

One of the ultimate goals of the theory of non-Markovian open systems is to contribute to better quantum technologies. A typical task in quantum information processing must eventually extend to larger multipartite systems consisting of $n$ qubits and therefore it is of great importance to understand the scaling properties of non-Markovianity measures.
Thus, as a final step to complete the picture on two-qubit non-Markovianity we study the relationship between the degrees of non-Markovianity of the two-qubit channel, denoted $\mathcal{N_{\text{2}}}$, and the local single-qubit channel denoted by $\mathcal{N_{\text{1}}}$.  

When a given measure requires optimization over initial states, the task of calculating non-Markovianty quickly becomes impractical for a larger number of qubits. It is therefore desirable to find links between the amount of non-Markovianity in multi-qubit channels and single-qubit channels. We present here a study of additivity of the BLP measure and show how this can vary greatly depending on the factorizability of the map.

When the two-qubit map factorizes in the limit of large qubit separation, $\Phi(t) = \Phi_A(t) \otimes \Phi_B(t)$, we discover a simple connection between the single-qubit non-Markovianity and the two-qubit non-Markovianity for the pure dephasing model:
\eq
\mathcal{N_{\text{2}}}=(e^{-\Gamma_0(b)}+e^{-\Gamma_0(a)})\mathcal{N_{\text{1}}}.
\eeq
Recall that $\Gamma_0(t)$ is the single-qubit decoherence factor, and the corresponding decay rate has at most, a single period of negativity in the ultracold realization of the dephasing model considered: $\gamma_1(t)<0$ when t $\in[a,b]$. Naturally for $a_B<a_B^{crit}$ we have $\mathcal{N}_1=0$ leading to $\mathcal{N}_2=0$: the two-qubit map is Markovian in the absence of single-qubit non-Markovianity and environment-mediated interactions. When $a_B>a_B^{crit}$, i.e., the single-qubit channel is non-Markovian, $e^{-\Gamma_0(b)}+e^{-\Gamma_0(a)}\lesssim2$, demonstrating that even in the simple case when the two-qubit channel factorizes into two identical single-qubit channels, the BLP measure is not additive. Instead, in this case, $\mathcal{N_{\text{2}}}< 2\,\mathcal{N_{\text{1}}}$, i.e., the measure is sub-additive.

Interestingly, in the case when the qubits share a common environment the situation is reversed and the BLP measure is superadditive, $\mathcal{N_{\text{2}}}> 2\,\mathcal{N_{\text{1}}}$. The additivity properties of the full dynamical map have been assessed numerically, and two instances are shown in Fig.~\ref{nM1}). The figures also show that the closer the qubits are to each other, the larger is the difference $|2\;\mathcal{N_{\text{1}}}-\mathcal{N_{\text{2}}}|$. This clearly demonstrates the ability of the shared environment to enhance memory-keeping properties of the two-qubit map.

Depending on whether one wants to enhance or suppress non-Markovianity in a register of qubits, the pure dephasing model can be engineered by decreasing or increasing the distance between qubits, respectively, allowing for an additional degree of control of the register dynamics.

It is an interesting open question if the sub-additivity of the BLP measure in the local environment scenario and super-additivity in the common environment case extend beyond the two-qubit case. An affirmative answer would imply that non-Markovianity scales more favorably in the common environment case. Understanding the additivity properties of non-Markovianity measures is significant when using memory as a resource and our results highlight the importance of a suitable physical realization of a qubit register: the scaling of the non-Markovianity measure can crucially depend on whether the qubits share the same environment or not.

\section{Conclusions}

We have studied a system of two qubits dephasing in a structured environment. By explicitly taking into account the spatial separation between the qubits we are able to interpolate between a common environment scenario, with the qubits close to each other, and an independent environment case, when the qubits are too far apart to influence each other. In the former case the dynamics is characterized by correlated noise, arising from environment mediated correlations. Our focus was on the degree of non-Markovianity of the two-qubit map, quantified either in terms of non-divisibility of the map, or by the amount of information that the system can recover from the environment. These two characterizations of non-Markovianity, which generally do not agree, were shown to coincide qualitatively for the model considered here.

Specifying our study to a physical realization of this model using optically trapped ultracold gases, we singled out the microscopic origin of memory effects. In this case non-Markovianity emerges from the interplay of correlated noise and single-qubit memory effects due to a non-trivial structure of the reservoir. When the local single-qubit dynamics is Markovian, two-qubit non-Markovianity can only emerge from correlated noise. More generally, the degree of non-Markovianity is amplified when the qubits share an environment, highlighting the key role of correlated noise in non-Markovian systems. Moreover, the effect of the common environment is also reflected in the global behavior of the non-Markovianity measured based on information flux. Actually, when the two qubits are very close to each other, the non-Markovianity measure is super-additive. Interestingly, when the qubits are in local environments the situation is reversed and the measure is sub-additive. Additivity properties affect the scaling of non-Markovianity measures for large qubit registers, potentially influencing the use of non-Markovianity as a resource for quantum technologies in a crucial way. We also discussed the pair of states optimizing the BLP measure and demonstrated the extreme sensitivity of the optimal pair on the system parameters.

\begin{acknowledgments}
This work was supported by the Engineering and Physical Sciences Research Council (Grant No. EP/J016349/1), the Scottish Doctoral Training Centre in Condensed Matter Physics,
the Emil Aaltonen foundation,the Magnus Ehrnrooth Foundation, and the Finnish Cultural
foundation. P. H. acknowledges A. Karlsson for useful discussions.
\end{acknowledgments}

\begin{appendix}
\section{Physical model}

In this work we consider a specific physical realization of the dephasing two-qubit model in order to explicitly explore the dynamics. The qubit takes physical form as an impurity atom occupying a double-well potential with an optical superlattice of wavelength $\lambda$. The distance between lattice sites, i.e., the size of the qubit, is $L = \lambda/4$. The minimum distance between qubits, where the two qubits occupy four consecutive lattice sites, is $2D_{min} = 8L$. The superlattice is immersed in a Bose-Einstein condensed (BEC) environment. We assume the condensate is trapped in a shallow potential and can be consider to be homogeneous. We also assume the gas to be weakly interacting, justifying the Bogoliubov approach. The specific forms of $\gamma_1(t)$ and $\gamma_2(t)$ for this system, originally derived in Ref.~\cite{originalmodel}, are

\begin{widetext}
\begin{eqnarray}
\label{G1}
\gamma_1(t)&=&\frac{g^2_{SE}n_0}{\hbar\pi^2}\int_0^{\infty}dk k^2e^{-k^2\sigma^2/2}\frac{\sin(\frac{E_k}{2\hbar}t)\cos(\frac{E_k}{2\hbar}t)}
{(\epsilon_k+2g_En_0)} \left(1-\frac{\sin(2kL)}{2kL}\right),\nn\\
\gamma_2(t)&=&\frac{g^2_{SE}n_0}{2\hbar\pi^2}\int_0^{\infty}dk k^2e^{-k^2\sigma^2/2}
\frac{\sin(\frac{E_k}{2\hbar}t)\cos(\frac{E_k}{2\hbar}t)}{(\epsilon_k+2g_En_0)}\left(\frac{\sin(2k(D+L))}{2k(D+L)}+\frac{\sin(2k(D-L))}{2k(D-L)}-2\frac{\sin(2kD)}{2kD}\right),\nn
\end{eqnarray}
\end{widetext}
	
where $g_E=4 \pi \hbar^2 a_E/m_E$ is the boson-boson coupling constant for a BEC environment with scattering length $a_E$ and mass $m_E$, and $g_{SE}=2\pi \hbar^2 a_{SE}/m_{SE}$ is the coupling between the condensate (environment) and the impurity atom (system) with scattering length $a_{SE}$ and reduced mass $m_{SE}=m_S m_E / (m_S + m_E)$. $E_k= \sqrt{2 \epsilon_k n_0 g_E + \epsilon_k^2}$ is the energy of the $k$-th Bogoliubov mode, where $n_0$ is the condensate density and $\epsilon_k= \hbar^2 k^2 / (2 m_E)$. Finally, $\sigma$ is the variance parameter of the lattice site. We consider specifically $^{23}$Na impurity atoms immersed in a $^{87}$Rb condensate, with $\lambda = 600nm$ and $n_0=10^{20}$m$^{-3}$. The natural scattering length of the rubidium atoms is $a_{Rb}=99 a_0$, where $a_0$ is the Bohr radius,  and we further assume $a_{SE}=55 a_0$. 

\end{appendix}


\end{document}